**Observation of Damped Oscillations in Chemical-Quantum-Magnetic Interactions**

Luana Hildever[1], Thiago Ferro[1] and José Holanda[1,2,*]

[1]Programa de Pós-Graduação em Engenharia Física, Universidade Federal Rural de Pernambuco, 54518-430, Cabo de Santo Agostinho, Pernambuco, Brazil

[2]Unidade Acadêmica do Cabo de Santo Agostinho, Universidade Federal Rural de Pernambuco, 54518-430, Cabo de Santo Agostinho, Pernambuco, Brazil

* Corresponding author: joseholanda.silvajunior@ufrpe.br
*Orcid iD: https://orcid.org/0000-0002-8823-368X

**Abstract**

Fundamental interactions are the basis of the most diverse phenomena in science that allow the dazzling of possible applications. In this work, we report a new interaction, which we call chemical-quantum-magnetic interaction. This interaction arises due to the difference in valence that the $Fe_3O_4$/PANI nanostructure acquires under certain conditions. In this study, PANI activates the chemical part of the oscillations, leaving the quantum and magnetic part for the double valence effect and consequently for changing the number of spins of the nanostructure sites. We also observed using interaction measurements that chemical-quantum-magnetic interactions oscillate in a subcritical regime satisfying the behavior of a damped harmonic oscillator.

**Introduction**

The effervescent area of spintronics has presented discoveries of new phenomena with enormous possibilities for applications [1-4]. Among the phenomena that stand out in these discoveries are the effects of spin Hall [5], Rashba-Edelstein [7], spin pumping [8], spin Seebeck [9], anomalous Hall [10], anomalous Nernst [11], spin Peltier [12], spin torque transfer [13], phonon spin detection [1], magnetic interfacial [3], the definition of magneto-optical states [15, 16, 17], the giant magnetoresistance [18], among many others [19, 20]. A common point of these effects is the interactions that originate them [16]. The fundamentals of these effects are quantum mechanics, the Pauli distribution, interactions of Heisenberg, Zeeman, Rashba, spin-orbit, chemical electronegativity, covalent bonds, double valence, and the definition of chemical potential, among many others [21, 22, 23]. All basic types of interactions have their biological, physical, or chemical origin. One of these interactions capable of attracting attention is from double valence, especially when it comes to transition metal ions such as iron. Iron is a chemical element that may have a valence of +2 or +3 and, under specific conditions, can even change its valence [24, 25, 26]. Among the materials that can present this change are iron oxide nanoparticles [25, 26, 27]. The nanoparticles have dimensions that are usually between 1 nm and 100 nm, which allow their use in geology, medical diagnosis, and improving the development of materials for data storage, among many other applications [27, 28]. Iron oxides are structures of Fe with O and (or) OH, and almost all are in the trivalent state ($Fe^{+3}$). Only three nanostructures contain a divalent state ($Fe^{+2}$), i.e., FeO, $Fe(OH)_2$, and $Fe_3O_4$. Furthermore, almost all iron oxides and hydroxide oxides are crystalline. However, the structural order and size of the crystallite can vary depending on the conditions under which the crystals are formed [24, 25, 26, 27, 28, 29, 30].

Three iron oxides stand out: magnetite, maghemite, and hematite. Magnetite ($Fe_3O_4$) has a face-centered cubic structure with a lattice parameter of $a = 0.839$ nm and a saturation magnetization on the order from 90 to 100 emu/g [29, 30]. Magnetite differs from other structures of iron oxides in that it contains divalent iron ions $Fe^{2+}$ and trivalent iron ions $Fe^{3+}$. Magnetite has an inverse spinel crystal structure, where its tetrahedral (purple) sites have ions $Fe^{3+}$, while the octahedral (purple/blue) site has both ions $Fe^{3+}$ (S = 5/2) and $Fe^{2+}$ (S = 2) [25, 26, 27, 28, 29, 30]. The arrays of the octahedron and tetrahedron, as well as the ordering of the spins in the magnetite unit cell, are shown in **Figure 1**, which presents that the spin of the eight ions $Fe^{3+}$ in the purple sites cancel with the eight ions $Fe^{3+}$ in the purple/blues sites. Therefore, the resulting magnetic moment is due exclusively to the $Fe^{2+}$ ions, which have spin S = 2. Magnetite has a Curie temperature of 850 K, and below this temperature, the spins at the purples and purple/blues sites align antiparallel. However, the magnitudes from the spins are different, making magnetite ferrimagnetic, whose spin array is $Fe^{3+}[Fe^{3+}+Fe^{2+}]O_4$. Maghemite ($\gamma Fe_2O_3$) has a similar structure to magnetite, such that it has an inverse spinel structure forming a face-centered cubic lattice with a lattice parameter of $a = 0.834$ nm and a saturation magnetization between 70 and 85 emu/g [29, 30 ]. Maghemite is the equivalent of oxidized magnetite, where during the process of oxidation, an ion $Fe^{2+}$ leaves the purple site, leaving a vacancy in the crystalline lattice, and another ion $Fe^{2+}$ transforms into ion $Fe^{3+}$; this is the main difference between maghemite and magnetite, that is, iron is in the trivalent state [31, 32, 33]. The formula for maghemite is $\gamma Fe_2O_3$ but may be re-written by multiplying it by 4/3, which results in $Fe_{8/3}O_4$, which emphasizes the similarity between the structure of maghemite and of magnetite [25, 26, 27, 28]. Regarding temperature, maghemite is ferrimagnetic at room temperature and transforms into hematite ($\alpha Fe_2O_3$) at temperatures above 800 K. In this context, hematite has a hexagonal unit cell with lattice parameters $a$ = 0.503 nm and $c$ = 1.375 nm and has a saturation magnetization of the order from 0.2 to 0.4 emu/g [25, 26]. Furthermore, at room temperature, it becomes weakly ferromagnetic, at 260 K (Morin temperature), undergoes a phase transition to an antiferromagnetic state, and above 956 K is paramagnetic [25, 26].

The structure of polymers consists of the repetition of small monomeric units connected through covalent bonds, resulting in large molecular chains, whose properties depend on the nature of the monomeric units, chain size, and crystallinity, among other properties [34, 35]. Polyaniline (PANI) has received attention due to its ease of synthesis, low cost, chemical stability, and high conductivity compared to other conductive polymers [34, 35]. For the preparation of PANI, the most used methods are chemical and electrochemical, the latter consisting of the oxidation of the aniline monomer ($C_6H_5NH_2$) in acidic electrolyte solutions such as $H_2SO_4$, HCl, and $HNO_3$, among many others. Regardless of the synthesis process, the structure of polyaniline in base form (undoped) has three fundamental states: oxidized, reduced, and oxidized-reduced. A completely oxidized state is called pernigranilin; a completely reduced state is called leucoeseraldine; and an oxidized-reduced state with 50% oxidation and 50% emeraldine base is called an intermediate state [36 ]. These three states are all insulating. However, to form the emeraldine salt, which is the conducting form of PANI, the emeraldine base reacts with the acid, producing protonation of the nitrogen atoms that bind to the benzenoid and quinoid rings, which separate to form polarons that will rise to as expected [34, 35]. The first PANI structure containing $Fe_3O_4$ nanoparticles was synthesized by mixing an aqueous $Fe_3O_4$ solution with an emeraldine base [36], demonstrating that the superparamagnetic properties observed in the structure are due to the $Fe_3O_4$ nanoparticles incorporated into PANI. Furthermore, we found that the properties of these systems (polyaniline and iron oxides) depend directly on particle size, interactions, and temperature. Due to this, several research groups began to synthesize these composites by modifying the synthesis routes to study the magnetic and conductive characteristics [34, 35, 36].

In the last five years, nanostructures containing iron oxide and PANI have been studied extensively due to the ideal combination of magnetic properties and polymeric conductivity [34, 35, 36]. More recent studies show that the $Fe_3O_4$/PANI structure has interesting electrical and magnetic properties, which one can use in applications such as data storage, ferrofluids, optoelectronics, spintronics, biomedical applications such as drug administration, degradation of dyes in the textile industry, among many others [24, 25, 26, 34, 35, 36]. In this work, we studied the fundamental interactions in $Fe_3O_4$ nanoparticles with polyaniline (PANI), which revealed that the magnetizing and demagnetizing interactions are related to the chemical-quantum interactions of the +2 and +3 valences of iron. The relationship between these interactions gave rise to a new interaction, which we call chemical-quantum-magnetic interaction, which presents an oscillating behavior in a damped regime.

**Experimental Section**

To produce our samples, we used commercial iron oxide nanoparticles from Sigma Aldrich and ultra-violet (UV) light from a Cole-Parmer 97620-42 lamp, which emits pure UV radiation with a wavelength of 365 nm. We performed chemical-quantum-magnetic interaction measurements using the Vibrating Sample Magnetometry (VSM) method, one of the most used systems to study magnetic properties [37]. With the VSM technique, it is mainly measured isothermal remanent magnetization curves (IRM (H)) and direct current demagnetization (DCD (H)) [15, 16, 17]. In an IRM(H) measurement, the starting point is the demagnetized sample after cooling in a zero magnetic field. Thus, a small magnetic field is applied after a time interval sufficient for magnetic equilibrium. Then, the small magnetic field is turned off, and the remanence is measured. The process is repeated, increasing the value of the magnetic field until the sample reaches saturation and the remanence assumes its highest value. The DCD(H) measurement is similar to IMR(H). However, the sample is initially in a saturated state. Then, the applied magnetic field is slowly reversed until a small value opposite the initial magnetization is identified. Afterward, the applied magnetic field is turned off and the remanence is measured. The process is repeated step by step, increasing the value of the inverted magnetic field until saturation. The purpose of these processes is to determine the Δm (H) curves [15, 16, 17].

**Theoretical section**

As presented in the experimental apparatus section, the Δm (H) curves are comparisons between the IRM(H) and DCD(H) curves, which are measured experimentally in the same way, differing only in the different magnetic states that the sample is in when is initiated each type of measurement. The proposal of Stoner and Wohlfarth [38, 39] demonstrates that there is a relation between IRM(H) and DCD(H) for non-interacting particles,

$$m_{d1}(H) = 1 - 2m_r(H), \tag{1}$$

where $m_{d1}(H) = DCD(H)_1/IRM(H_{Max})$ and $m_r(H) = IRM(H)/IRM(H_{Max})$. Visualizing different behavior, Henkel proposed that the deviation of this behavior in a real system was due to the interactions between the nanoparticles [40]. In systems with magnetic interaction, the experimental data distances itself from the curve made with equation (1). Thus, this is notable since the equation (1) assumes that the magnetizing and demagnetizing processes are equivalent. Qualitatively the type of interaction was defined by the introduction of the term Δm (H) in equation (1) [15, 16, 17],

$$m_{d2}(H) = \Delta m(H) + [1 - 2m_r(H)], \tag{2}$$

where $m_{d2} = DCD(H)_2/IRM(H_{Max})$. For Δm (H) < 0 the predominant interactions are demagnetizing (PID) and if Δm (H) > 0 the predominant interactions are magnetizing (PIM) [15, 16, 17]. In **Figure 1**, we show a schematic of the oscillatory behavior due to the chemical-quantum-magnetic interaction between the two ions iron with different valences and number of spins. The system works as a type of damped harmonic oscillator, which is one of the best-known physical systems in physics, and for the present case, obeys the following equation:

$$\frac{d^2\Delta m(H)}{dH^2} + \frac{1}{H_D}\frac{d\Delta m(H)}{dH} + \frac{1}{H_M^2}[\Delta m(H) - \Delta m(H=0)] = 0. \tag{3}$$

The H$_D$ and H$_M$ fields are the demagnetizing and magnetizing fields, respectively [15, 16, 17]. The demagnetizing and magnetizing fields are associated with the dissipation and gain of magnetic energy, respectively, and both are due to the chemical-quantum-magnetic interaction.

**Results**

The double valence interaction showing the change in the number of spins of the ions determines the chemical-quantum interaction, such that the magnetic contribution due to the ions can increase the local magnetism of the nanostructure. This behavior underlies our observation of a new interaction, which we call chemical-quantum-magnetic interaction. This interaction between ions contributes to the magnetic flux generating oscillations in this interaction. **Figure 1** shows a schematic of the characteristics that caused the chemical-quantum-magnetic interactions to oscillate in the Fe$_3$O$_4$/PANI nanostructures.

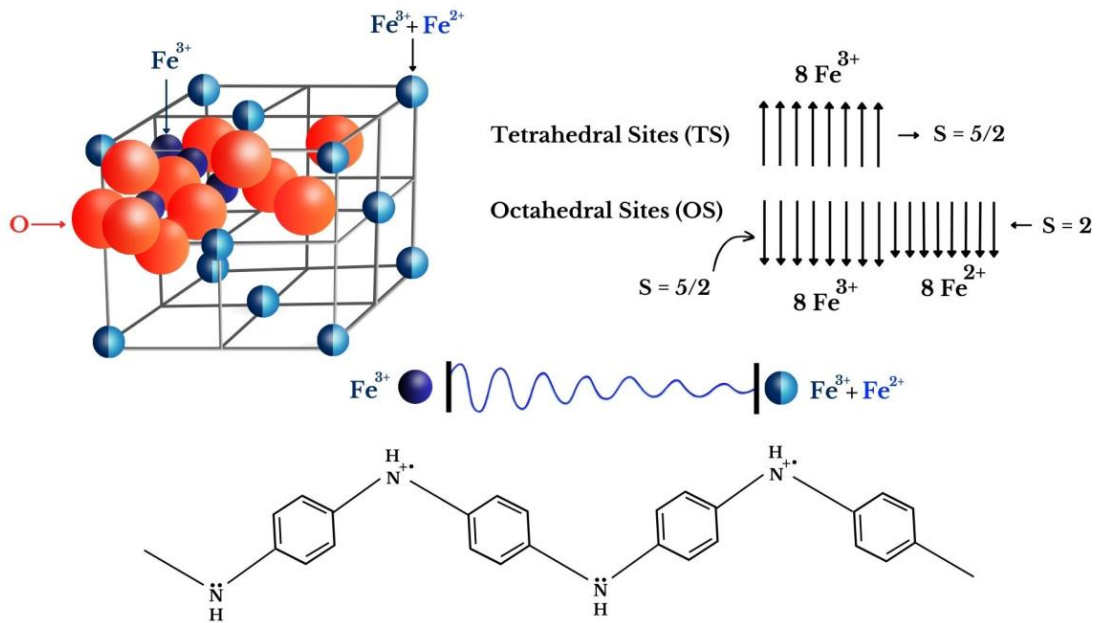

**Figure 1.** Schematic showing that magnetite has an inverse spinel crystal structure with tetrahedral sites (purple) occupied by ions Fe$^{3+}$ and octahedral sites (purple/blues) occupied by both ions Fe$^{3+}$ (S = 5/2) and Fe$^{2+}$ (S = 2) [25, 26, 27, 28, 29, 30]. Furthermore, it shows that the chemical-quantum-magnetic interaction produces oscillations as in a damped harmonic oscillator.

The behavior that these interactions can acquire depends on the manufacturing process of the nanostructures [15, 16, 17]. In **Figure 2**, we present measurements of Δm (H) curves performed on a group of $Fe_3O_4$/PANI nanostructures synthesized at a temperature of 310 K using the methodology presented in the experimental section. The time t in **Figure 2** represents the exposure time of the nanostructures to UV radiation before Δm (H) measurements. The results in **Figures 2 (a)-(g)** reveal that only the PID state is present at this synthesis temperature; this way, there is no net gain in magnetic energy. Thus, it is not possible to observe oscillations due to the change in valence, that is, due to chemical-quantum-magnetic interaction. This approach is in accord with Eq. (3) because if the third term of this equation does not exist, it only reveals the fact that chemical-quantum-magnetic interactions are proportional to the variation of the magnetic field, which is in accord with the models of Stoner-Wohlfarth [38, 39], Henkel [40], J. Holanda [15, 16, 17] and with the result presented in **Figure 2 (h)** for the intensities of the interactions as a function of the time of synthesis t. We obtained the intensity values from the interactions presented in **Figure 2 (h)**, considering the data of **Figures 2 (a)-(g)** and the models from references [15], [16], [17], [38], [39] and [40]. Furthermore, for t = 60 minutes (min.), we find the maximum intensity of the chemical-quantum-magnetic interactions of the PID state.

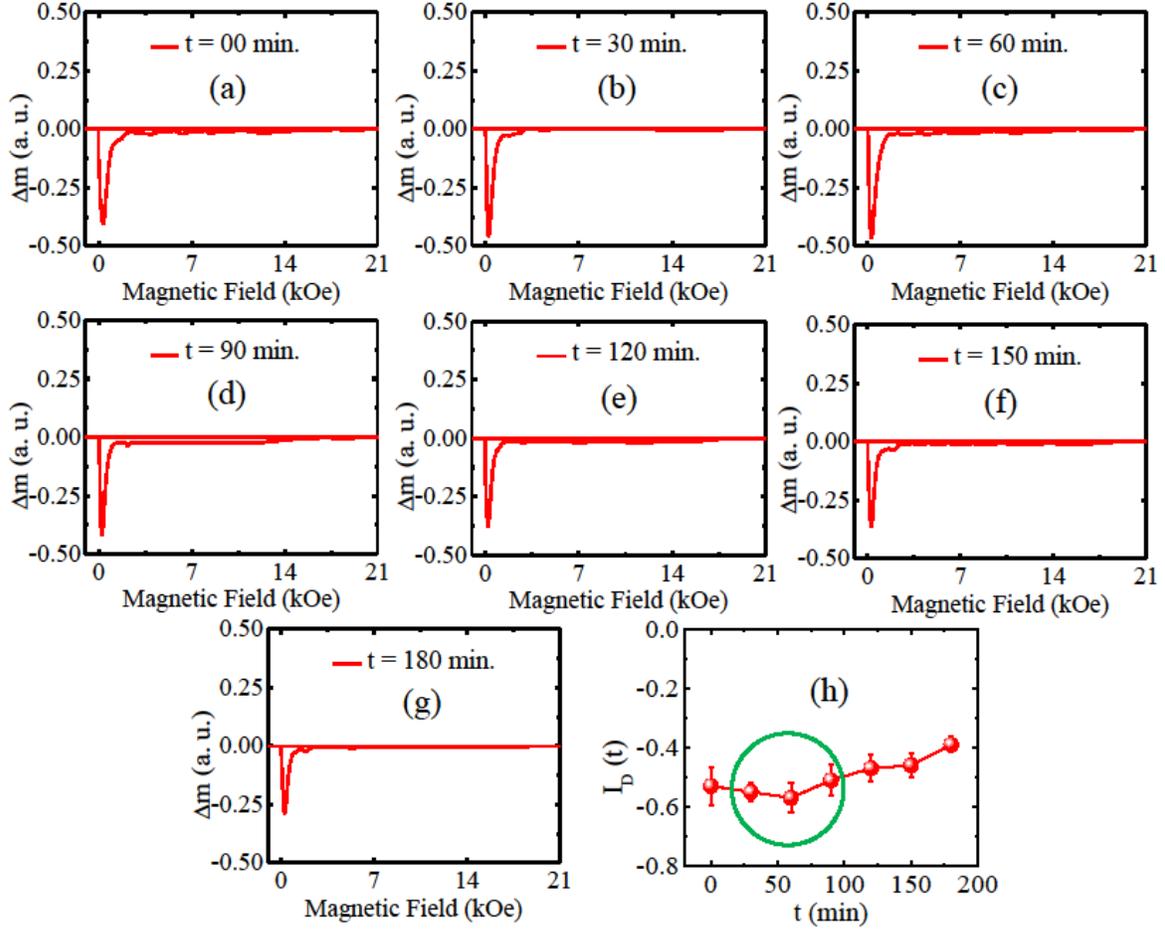

**Figure 2.** Shows measurements of Δm (H) curves performed on Fe$_3$O$_4$/PANI nanostructures synthesized at a temperature of 310 K using the methodology presented in the experimental section. Time t represents the exposure time of the nanostructure to UV radiation before Δm (H) measurements. **(a)** t = 0 min., **(b)** t = 30 min., **(c)** t = 60 min., **(d)** t = 90 min., **(e)** t = 120 min., **(f)** t = 150 min., **(g)** t = 180 min., and **(h)** intensities of interactions as a function of synthesis time t for the PID state of chemical-quantum-magnetic interactions.

Our second group of samples was synthesized at a temperature of 330 K, where we also varied the time t of exposure to UV radiation, as shown in **Figure 3**. For the conditions presented, we observed large oscillations in the chemical-quantum-magnetic interactions with a maximum amplitude for the time from t = 60 min. Such an oscillation effect has the contributions of dissipation and gain of magnetic energy and agrees with Eq. (3). We also note that the amplitude of the oscillations is sensitive to the time of synthesis t, that is, to the exposure time to UV radiation, as shown in **Figures 3 (a) – (g)**. **Figures 3 (h)** and **(i)** confirm that the ideal conditions for obtaining maximum oscillations are for a time of t = 60 minutes. Furthermore, **Figures 3 (h)** and **(i)** separately show the intensities of the interactions of the PID and PIM states, respectively. A notable behavior of this oscillation process is its evolution towards stability, as shown in **Figure 3 (g)**, which unequivocally highlights the energetic balance between dissipation and gain of magnetic energy. As far as we know, this is the first evidence of this phenomenon.

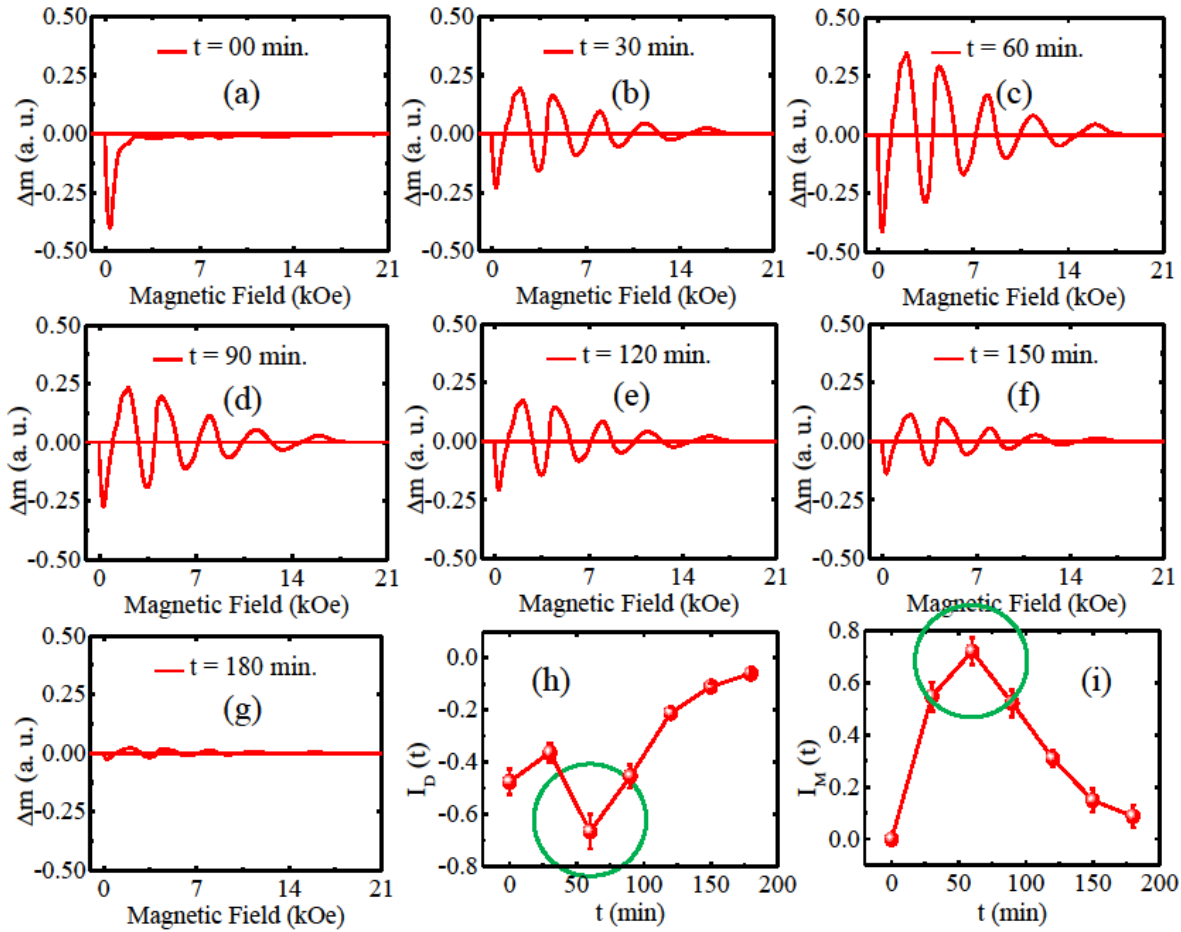

**Figure 3.** Δm(H) curves with large oscillations due to the chemical-quantum-magnetic interactions for the $Fe_3O_4$/PANI nanostructures synthesized at a temperature of 330 K according to the methodology presented in the experimental section. Time t represents the exposure time of the nanostructures to UV radiation before Δm (H) measurements. **(a)** t = 0 min., **(b)** t = 30 min., **(c)** t = 60 min., **(d)** t = 90 min., **(e)** t = 120 min., **(f)** t = 150 min., **(g)** t = 180 min., **(h)** and **(i)**

intensities of interactions as a function of time of synthesis t for the PID and PIM states, respectively.

To have a direct comparison of the results presented in **Figure 3** and equation (3), we analytically solved equation (3) considering particularly the subcritical solution, which is the solution that best describes our observed oscillations, that is,

$$\Delta m(H) = \Delta m(H=0) + \Delta \widetilde{m}(H) e^{-H/H_D} \cos\left[\left(\frac{H}{H_{eff}} + \varphi_{DM}\right)\right] \qquad (4)$$

with

$$\frac{1}{H_{eff}^2} = \frac{1}{H_M^2} - \frac{1}{H_D^2}. \qquad (5)$$

**Figure 4** presents the results obtained with equations (4) and (5). The parameters used are in **Table 1** for t = 60 and 120 min. The terms $\Delta m_{(+)}(H) = \Delta m(H, \Delta \widetilde{m}(H) > 0)$ and $\Delta m_{(-)}(H) = \Delta m(H, \Delta \widetilde{m}(H) < 0)$ define the amplitudes of the subcritical state of the oscillations, which represent the blue lines in **Figure 4**. In a more general context, oscillations open a new direction for research in nanostructures with ferromagnetic and polymeric materials since they are similar to those found in spintronic nano-oscillators, chaotic systems, nonlinear electronic circuits, and even in gravitational waves [41, 42, 43].

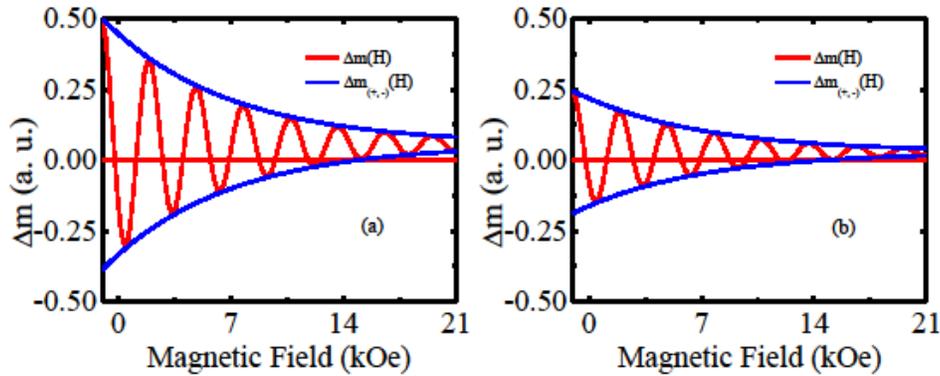

**Figure 4.** Presents the results obtained with equations (4) and (5) using the parameters shown in **Table 1** for **(a)** t = 60 min., and **(b)** 120 min. The terms $\Delta m_{(+)}(H) = \Delta m(H, \Delta \widetilde{m}(H) > 0)$ and $\Delta m_{(-)}(H) = \Delta m(H, \Delta \widetilde{m}(H) < 0)$ define the amplitudes of the subcritical state of the oscillations, which represent the blue lines.

**Table 1.** Parameters used in equations (4) and (5) to obtain the results presented in **Figure 4** for t = 60 and 120 min..

| t (min.) | $\Delta m$ (H = 0) (a. u.) | $\Delta \tilde{m}$ (H) (a. u.) | $H_D$ (Oe) | $H_M$ (Oe) | $\varphi_{DM}$ (°) |
|---|---|---|---|---|---|
| 60 | 0.056 | ±0.39 | 7.7 | 0.008 | -56.5 |
| 120 | 0.028 | ±0.19 | 7.7 | 0.008 | -56.5 |

To emphasize that our results are relevant, we show in **Figure 5** measurements of $\Delta m$ (H) curves in another set of samples synthesized at a temperature of 350 K, where we also varied the exposure time t. We observed the same behavior related to **Figure 2** for this set of samples, which reveals that the synthesis temperature of the samples contributes to activating damped oscillations in chemical-quantum-magnetic interactions. The effect has a narrow range of activation of observed behavior, similar to the oscillations in spintronic nano-oscillators produced by the anti-damping effect [44]. In order of quantum chemical thermodynamics, the oscillations represent fluctuations proportional to the synthesis temperature and are sensitive to the magnetic behavior of the ions in the nanostructure array [45]. Furthermore, **Figures 5(a)-(g)** also show that only the PID state is well-defined for this synthesis temperature, as evidenced in **Figure 5(h)**, which presents a pure linear behavior for $\Delta m$ (H ) as a function of t, which is not so evident in **Figure 2(h)**. This behavior shows that the temperature range for the chemical-quantum-magnetic interaction to oscillate is less than 350 K.

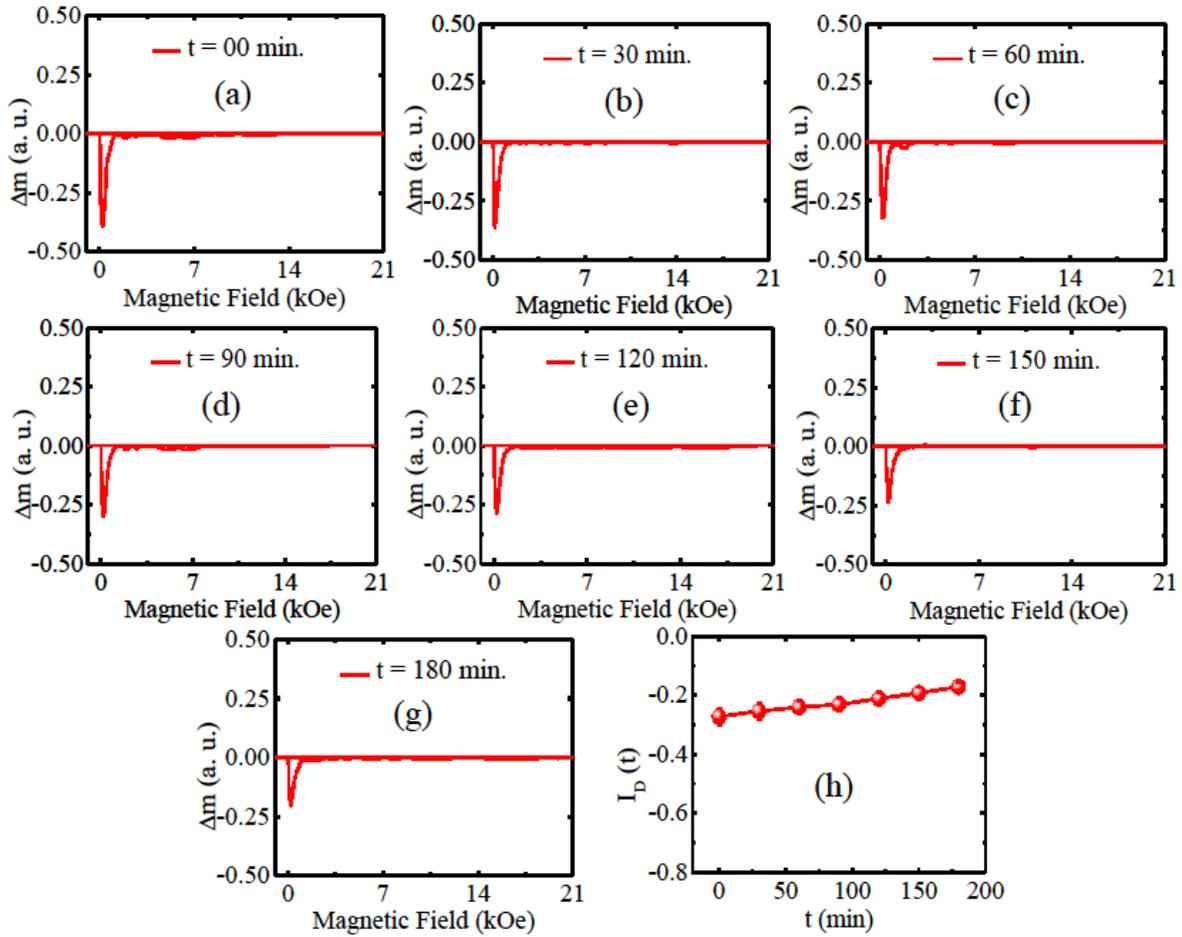

**Figure 5.** Shows measurements of Δm (H) curves performed on $Fe_3O_4$/PANI nanostructures synthesized at a temperature of 350 K using the methodology presented in the experimental section. Time t represents the exposure time of the nanostructures to UV radiation before Δm (H) measurements. **(a)** t = 0 min., **(b)** t = 30 min., **(c)** t = 60 min., **(d)** t = 90 min., **(e)** t = 120 min., **(f)** t = 150 min., **(g)** t = 180 min., and **(h)** intensities of interactions as a function of synthesis time t for the PID state of chemical-quantum-magnetic interactions.

The behaviors presented in this work reveal the existence of a new fundamental interaction, which we call chemical-quantum-magnetic interaction. Both the synthesis process and exposure to UV radiation influence the activation of oscillations, causing ions with different valences to oscillate. As the DCD(H) and IRM(H) curves obey thermodynamic cycles with magnetic field variation, the oscillatory behavior is sensitive to the magnetic field, allowing its amplitude.

**Conclusion**

The physical entities of exceptional interest in any area are interactions, as they are responsible for different effects that may arise from them. In this work, we report a new fundamental interaction called chemical-quantum-magnetic, which appears due to the difference in valence that the $Fe_3O_4$/PANI nanostructure acquires under certain conditions. In this study, PANI activates the chemical part of the oscillations, leaving the quantum and magnetic part for the double valence effect and consequently for changing the number of spins of the nanostructure sites. Our studies are pioneering in the topic while finding similarities in systems such as spintronic nano-oscillators produced by the anti-damping effect. Regarding our groups of samples, we noticed that the synthesis temperature range in which the oscillations appear is around the temperature from $330 \pm 5$ K for $Fe_3O_4$/PANI nanostructures with an average diameter of $44 \pm 1.2$ nm. The interpretation of our experimental data with a damped harmonic oscillator-type equation highlights the naturalness of the observed phenomenon. In summary, we are convinced that our studies will be sources of future studies using this new type of interaction and its oscillations.

**Acknowledgements**

This research was supported by Conselho Nacional de Desenvolvimento Científico e Tecnológico (CNPq), Coordenação de Aperfeiçoamento de Pessoal de Nível Superior (CAPES), Financiadora de Estudos e Projetos (FINEP), and Fundação de Amparo à Ciência e Tecnologia do Estado de Pernambuco (FACEPE). The authors are grateful to prof. Changjiang Liu of Physics Department of The State University of New York at Buffalo (University at Buffalo) by the discussions regarding sample preparation, and to prof. F. Chávez for help with the experimental measures, which were financially supported through the projects: processes No.: IBPG-1292-3.03/22/FACEPE/UFRPE, No.: 23082.015969/2022-31 UACSA/UFRPE, No. 23082.012820/2022-09 PPGENGFIS/UFRPE, No.: 309982/2021-9 CNPq/UFRPE, No.: 23082.025190/2022-24 CAPES/PRPG/UFRPE and proposal No. PSTE: PS0079/23-0001 approved at the Centro de Tecnologias Estratégicas do Nordeste (CETENE), which is a Research Unit of the Ministry of Science, Technology and Innovation (MCTI) from Brazil.

**Contributions**

L. H. and T. F. analyzed all the experimental measures and J. H. discussed, wrote and supervised the work.

**Conflict of interest**

The authors declare that they have no conflict of interest.